\documentclass[runningheads]{llncs}

\usepackage[T1]{fontenc}
\usepackage{graphicx}
\usepackage{amsmath}
\usepackage{amssymb}
\usepackage{booktabs}
\usepackage{array}
\usepackage{hyperref}
\usepackage{url}
\usepackage{microtype}
\usepackage{xcolor}
\usepackage{float}
\usepackage{orcidlink}
\usepackage{tikz}
\usetikzlibrary{positioning, arrows.meta, shapes.geometric, fit, backgrounds, calc}

\begin{document}

\title{Conflicted Records: Typed Adversarial Contracts\\
for Auditable LLM Output Quality Enforcement}

\titlerunning{Conflicted Records: Typed Adversarial Contracts for LLM Quality}

\author{Kardelen Subasi\orcidlink{0009-0005-5073-4656} \and
  Jaros\l{}aw A. Chudziak\orcidlink{0000-0003-4534-8652}}
\authorrunning{Subasi and Chudziak}
\institute{Warsaw University of Technology, Warsaw, Poland\\
  \email{\{kardelen.subasi.stud, jaroslaw.chudziak\}@pw.edu.pl}}

\maketitle

\begin{abstract}
LLM-generated records increasingly populate production information systems
where they are treated as data assets subject to the same quality
expectations as records from operational databases. A single such record
can simultaneously satisfy and violate a quality rule: a grounded fact
adjacent to a hallucination, a useful answer adjacent to a stereotype.
Classical data cleaning cannot resolve this because the two judgments are
incompatible over the same span; a single LLM judge cannot either, since
it shares the training biases that produced the conflict. Existing
interventions such as prompt engineering, chain-of-thought, and RLHF
reduce surface error rates but provide no audit trail and cannot
distinguish genuine remediation from blanket suppression. The question is
whether a deterministic check over structured inter-agent records can catch
the defects that referential checks alone miss.

We present BiasGuard-D: two adversarial roles drawn from distinct model
families exchange typed claim documents, a three-model quorum adjudicates
by majority vote, and a combined referential-plus-semantic invariant
validates the verdict. The semantic component fires on verdicts referential
checks miss; removing the iterative loop drops invariant satisfaction
sharply. The deliberation layer transfers from stereotype to hallucination
remediation on HaluEval QA with only role-framing changes, raising
right-answer recovery by 49 percentage points; the triage gate required a
cross-family second auditor to maintain recall on objective hallucination.
Every claim, vote, and instruction is stored as a typed record giving
operators a provenance trail auditable under emerging AI regulations.
\end{abstract}


\section{Introduction}
\label{sec:introduction}

LLM outputs are no longer transient. They are persisted into search indexes,
written into knowledge bases, embedded in recommendation feeds, and surfaced
through customer-facing information systems where they act as data~\cite{lewis2020rag,edge2024graphrag}.
As deployments scale into regulated domains, operators need more than a quality
score; they need to know \emph{why} an output passes or fails a constraint,
\emph{which span} triggered the verdict, and \emph{whether the remediation
instructions are internally consistent}.

What makes this hard is that an LLM-generated record can satisfy and violate
a quality rule in the same sequence of text. A hiring recommendation may
accurately ground its reasoning in a candidate's qualifications while
simultaneously making an age generalisation the input does not support. A
knowledge-base entry may pair a faithfully lifted fact with one the model
invented. In neither case are the two things in separate fields or even
separate sentences; they share the same characters. Classical data
cleaning, which reads a record and refines it, cannot resolve this, because
the two issues require opposite judgments about the same text.
A single LLM judge cannot either, since it shares the same training biases
that produced the conflict~\cite{kim2025correlatederrors,zhang2025stopovervaluing,zheng2023judging}.

The question this paper asks is: \emph{can a deterministic check over
structured inter-agent records catch the quality defects that referential
checks alone miss, and do those records carry enough information to audit
pipeline quality without calling an additional model?} We evaluate this across
three dimensions: whether a semantic check catches a recurring class of
contradictions that referential checks miss; whether stored records carry
enough structure to tell genuine resolution from suppression without calling
a further model; and whether the deliberation layer transfers across different
quality domains with only role-framing changes.

We address this by treating enforcement as a data-quality problem rather than
a single-judge problem: two adversarial roles with opposing biases produce
structured claims about specific text spans; a multi-family quorum adjudicates;
and a deterministic invariant validates the result before any downstream action.
Records with a consistent verdict pass with a full provenance trail;
contradictory verdicts are quarantined rather than auto-repaired, because
auto-repair would bring back the model judgment the architecture is designed
to avoid. The experimental study confirms all three evaluation questions and
surfaces a finding we did not anticipate: the iterative deliberation loop
matters far more than any individual model choice, and the audit trail is
what makes this visible.

This paper makes three contributions. We present a five-stage typed
adversarial contract pipeline in which two adversarial roles from distinct
model families exchange typed claim documents over bounded peer-to-peer rounds,
a three-model quorum adjudicates, and a combined deterministic invariant
($\mathcal{I}_{\mathrm{ref}} \wedge \mathcal{I}_{\mathrm{sem}}$) validates
each verdict, making the full lineage from any instruction back to its
originating span queryable as data. An ablation shows that the iterative
deliberation loop accounts for most of the invariant-satisfaction gain:
removing the loop drops satisfaction by nearly 30 percentage points on the
BBQ Disability benchmark, while schema-level coordination metrics stored as
typed records make the degradation auditable without calling an additional
model. Finally, we provide cross-domain portability evidence showing that the
deliberation layer, data model, and invariant transfer from subjective
stereotype remediation to objective hallucination remediation by changing only
role-framing prompts; the triage gate required a cross-family second auditor
to maintain recall on objective hallucination, isolating the gate as the one
domain-sensitive component.

\begin{figure}[!ht]
\centering
\includegraphics[width=0.82\textwidth]{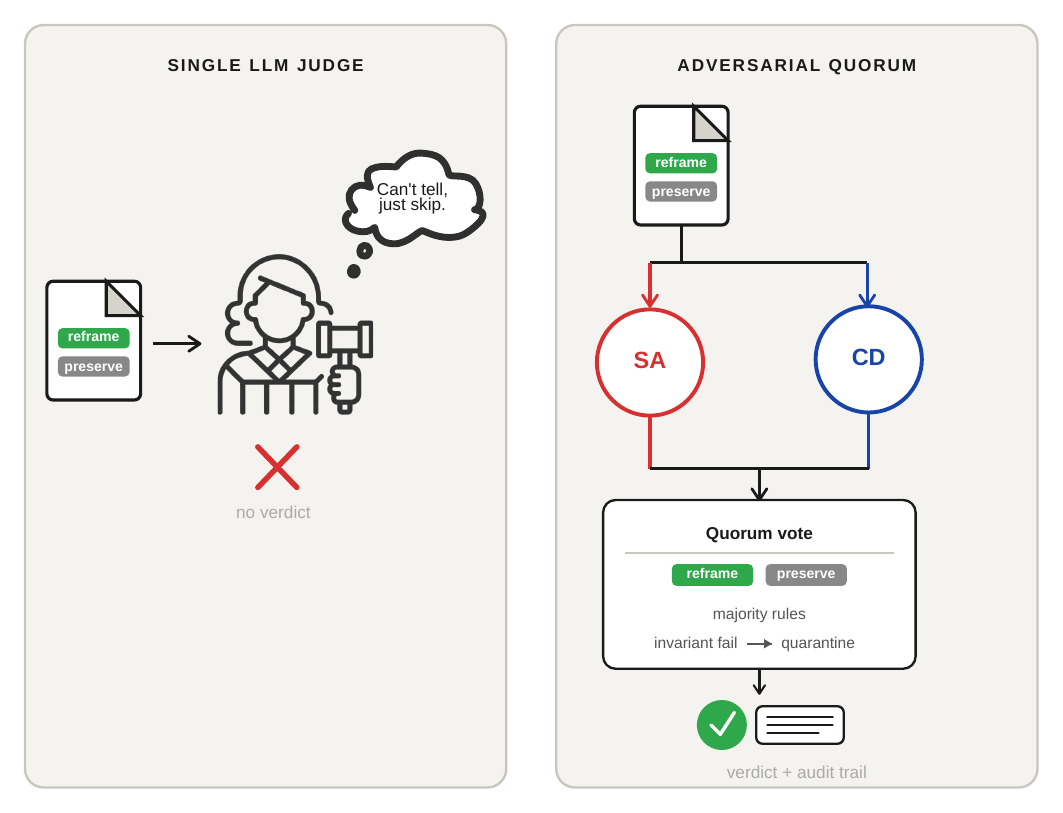}
\caption{A conflicted record (input): co-located spans require opposite
  actions (one contains demographic reasoning that must be reframed or dropped;
  another is contextually grounded and must be preserved) that a single LLM
  judge cannot resolve consistently and returns no verdict. BiasGuard-D returns
  a typed per-span instruction record (output) with a full provenance trail.}
\label{fig:io}
\end{figure}

Section~\ref{sec:background} reviews related work and shows where existing
approaches fall short. Section~\ref{sec:core} defines the problem and describes
the data model, pipeline, and invariant. Section~\ref{sec:evaluation} evaluates
the system on stereotype and hallucination benchmarks;
Sections~\ref{sec:discussion} and~\ref{sec:conclusion} discuss the findings,
limitations, and what the provenance trail offers in practice. Source code,
configuration, and run artefacts are at \url{https://github.com/ksubasi/biasguard-d}.

\section{Background and Related Work}
\label{sec:background}

BiasGuard-D sits at the intersection of classical data quality, LLM output
enforcement, and typed-contract agent orchestration. Each tradition contributes
something the system needs; none alone addresses the conflicted-record problem.

\paragraph{Constraint-based data quality and repair.}
The standard view of data quality treats records as entities that satisfy or
violate integrity constraints, and routing or repairing violators is a
long-studied problem~\cite{fan2012dependencies,ilyas2019datarepair,bertossi2011repairs,chomicki2007cqa}.
Our invariant is recognisable in this tradition: $\mathcal{I}_{\mathrm{ref}}$
is a foreign-key coverage check, and $\mathcal{I}_{\mathrm{sem}}$ resembles
near-duplicate detection in entity resolution. Classical repair theory selects
a minimal change to restore integrity, but a span instructed both to be removed
and preserved cannot be fixed this way; there is no ordering on
natural-language paraphrases that singles out a canonical repair.
BiasGuard-D quarantines on $\mathcal{I}_{\mathrm{sem}}$ failure rather than
attempting an auto-repair, because doing so would bring model judgment back
into the verification step. The novelty lies not in either predicate alone but
in applying them together over natural-language payloads.

\paragraph{Quality enforcement on LLM outputs.}
Retrieval-augmented and self-refining pipelines~\cite{lewis2020rag,edge2024graphrag,asai2023selfrag}
read, refine, and pass on records. This works well for straightforward quality
issues but fails on conflicted records because no stage can mark two co-located
spans as simultaneously valid and invalid. Recent multi-agent
debate~\cite{du2023improving,liang2023encouraging} introduces structured
disagreement but without typed message formats or deterministic verification.
Related work in the group has studied agent disagreement as a diagnostic signal
for moderation~\cite{wawer-agents-disagree-2026}, applied dialectical
multi-agent refinement to structured classification~\cite{baba-dialectical-2026},
and used community-driven agent frameworks for implicit bias
detection~\cite{gajewska-hatespeech-2026}.
Training-time approaches (Constitutional AI~\cite{bai2022constitutional} and
RLHF~\cite{ouyang2022instructgpt}) embed quality constraints into model
parameters; the resulting guarantees are statistical and opaque to an auditor.
A single-model pipeline that produces a cleaner-looking output provides no
evidence that the conflict was resolved rather than suppressed.
BiasGuard-D is runtime, record-level, and inspectable: every inter-stage
exchange is a typed record rather than a raw string.

\paragraph{Typed adversarial contracts and provenance.}
Typed inter-agent messaging descends from FIPA contract-net
protocols~\cite{fipa2002contractnet,smith1980contractnet}, was revived in
LLM orchestration frameworks~\cite{wu2023autogen,langgraph2024}, and has
acquired formal treatment in Hoare-style contracts~\cite{dewes2025contracts}.
Because every inter-stage record conforms to a fixed schema, the full lineage
from any instruction back to its source is stored and queryable without
re-running the models. Prior work supplies the message-passing machinery but
does not apply it to adversarial quality enforcement with a deterministic
check across different constraint types.

\section{Problem, Approach, and Methods}
\label{sec:core}

\subsection{The Problem}
\label{sec:problem}

Making the problem precise requires moving from record-level pass/fail to
span-level decomposition. A record that must simultaneously preserve one span
and remove another cannot be handled as a unit: each span needs an identified
role, and the roles assigned by different judges need to be checked for
mutual consistency before any downstream action.

Concretely: an operator maintaining an AI-generated catalogue needs both
span-level granularity to approve or remove individual claims without
discarding the whole entry, and a provenance trail linking each instruction
back to its source span, auditable without re-running the pipeline.

Both scenarios impose the same requirement on the data model: span-level
typed claims with stable identifiers, a deterministic consistency check
that needs no model at verification time, and a stored lineage that survives
the full pipeline. The data model described next is designed around exactly
these three requirements.

\subsection{The Data Model}
\label{sec:datamodel}

Every correctness property of BiasGuard-D rests on the data model, so we
describe it before the pipeline. Record-level pass/fail labels are
insufficient for quality assessment: locating where errors occur requires
finer-grained units~\cite{guerreiro-xcomet-2023,wang-factcheckbench-2024}.
Two design decisions follow from the problem.
First, fixing schemas up-front means stages can be swapped or upgraded without
ambiguity, which matters as models are deprecated. Second, grounding all
schemas in a single atomic primitive means the consistency invariant operates
on stable identifier sets rather than on natural-language content, keeping
verification deterministic and LLM-free.

The primitive is an \texttt{AtomicClaim}: one identifiable assertion about a
specific text span. Each claim carries a role-prefixed identifier
(\texttt{C001} for the Stereotype Advocate's first claim, \texttt{D001} for
the Context Defender's), the verbatim or compressed span it concerns, a
typed claim category drawn from five values (\textsc{Factual},
\textsc{Rhetorical}, \textsc{Framing}, \textsc{OptionPreference},
\textsc{DecisionFlip}), an optional list of provenance identifiers from a
reference passage, and a confidence score. The role prefix makes cross-role
provenance trivial to compute: finding where SA and CD claims overlap is a
string-prefix operation, not a semantic search.
Confidence is discounted multiplicatively each time the opposing role contests
a claim, so four contestations reduce it by roughly two-thirds rather than
collapsing it to zero; the trajectory is stored per round and flags cases where
a low-confidence claim is nevertheless upheld. Full field definitions and the
Python implementation are in the repository.

Everything else in the system is either an \texttt{AtomicClaim} or a typed
collection of them. Table~\ref{tab:schemas} lists the six inter-stage schemas.
The two primary documents (\texttt{StereotypeCase}, \texttt{ContextDefense})
are updated in place across rounds; the \texttt{DeliberationVerdict} carries
the output the invariant checks. Because every edge carries a typed record
with consistent identifiers, the full lineage of any instruction back to its
source span is recoverable by traversing stored records, with no
natural-language processing needed.

\begin{table}[!ht]
\caption{The six typed inter-stage messages. Every schema is either an
\texttt{AtomicClaim} or a collection of them.}
\label{tab:schemas}
\centering
\small
\begin{tabular}{p{33mm}p{20mm}p{60mm}}
\toprule
\textbf{Schema} & \textbf{Producer} & \textbf{Purpose and key fields} \\
\midrule
\texttt{VerdictContract}     & Auditor   & Triage: \texttt{delegate} flag, bias rationale, stereotype category \\
\texttt{StereotypeCase}      & SA        & SA's claim set: harmful \texttt{AtomicClaim} list, reframe targets, severity \\
\texttt{ContextDefense}      & CD        & CD's claim set: preserved \texttt{AtomicClaim} list, loss-risk score \\
\texttt{SecondRoundSA/CD}    & SA or CD  & Revision: contested/defended claim IDs, contestation grounds \\
\texttt{DeliberationVerdict} & Arbiter $\times 3$, then quorum & Adjudication: upheld SA/CD claims, remediation instructions (reframe / preserve / drop) \\
\bottomrule
\end{tabular}
\end{table}

\subsection{Pipeline Architecture}
\label{sec:architecture}

With the data model fixed, the pipeline determines which stage produces which
schema and how stages connect to preserve provenance end to end.
BiasGuard-D runs in five stages orchestrated by LangGraph~\cite{langgraph2024};
the orchestrator is interchangeable since every interface is typed, following
contract-based design principles for multi-agent
systems~\cite{dewes2025contracts}.

\paragraph{Generator and same-family Auditor (Stages 1--2).}
The Generator (Llama 3.1 8B via Groq) produces the incoming record from
structured input; the Auditor (same model, same family) emits a
\texttt{VerdictContract} carrying a \texttt{delegate} flag. Pairing the Auditor
with the Generator's own family is deliberate: self-preference bias means a
same-family judge is less likely to flag its own family's outputs~\cite{kim2025correlatederrors,zhang2025stopovervaluing}, so records
it does flag are ones where even a sympathetic reading finds a problem,
making it a high-precision, if lower-recall, triage gate. The deliberation pipeline is
invoked sparingly and only on records the same-family judge cannot explain
away; the recall cost is the precision-recall trade-off the architecture
already surfaces at the delegation threshold. Records where
\texttt{delegate=false} pass through unchanged; otherwise they enter
deliberation. For the cross-domain HaluEval condition a second cross-family
Auditor B (Mistral Small) is added; delegation is \texttt{OR(A, B)}, raising
recall at the cost of additional false delegations on correct inputs.

\paragraph{Bounded peer-to-peer deliberation (Stage 3).}
The Stereotype Advocate (SA, GLM-4-Flash via Zhipu AI) and the Context
Defender (CD, Mistral Small via Mistral AI) produce their initial
\texttt{StereotypeCase} and \texttt{ContextDefense} in round one. In each
subsequent round each role reads the other's previous document and re-emits
its own, reusing the prior round's \texttt{claim\_id} when a claim survives
semantically and assigning a fresh identifier when a substantively new claim
is raised. The stability check is deterministic and operates on identifier sets only: the
loop terminates when two consecutive rounds produce identical contested and
defended claim sets, or when a five-round cap is reached. The cap is set at
roughly twice the empirically observed mean convergence point (2.24--2.39
rounds across conditions; Table~\ref{tab:coordination}): if claim sets have not
stabilised by that point, continued exchange is unlikely to resolve the
underlying tension, and the invariant will surface the contradiction at Stage
5 regardless. The two roles update in parallel within a round, so
neither sees the other's current-round output. Identifier mis-assignment by a
role delays stability by at most one round; it does not compromise the invariant.

\paragraph{Quorum Arbiter and invariant check (Stages 4--5).}
Three Arbiters drawn from three distinct provider families (Cohere Command R,
Claude Sonnet, and Llama 3.3 70B) receive the converged claim documents
in parallel and each emit an independent \texttt{DeliberationVerdict}. A
per-claim majority vote (at least two of three) produces the aggregated verdict,
which then passes through the combined invariant (\S\ref{sec:invariant}). If
the invariant holds, the Generator is re-invoked with the upheld preserve,
reframe, and drop instructions; if either component fails, an
\texttt{InvariantViolation} record is logged and the pipeline emits a
conservative fallback. The quarantine path deliberately avoids auto-repair: a
verdict that fails the semantic check has gone through five stages of
deliberation without resolving its contradiction, and patching it over would
bring model judgment back into the verification step.

\paragraph{Family heterogeneity as a data-integrity guarantee.}
The five roles span five distinct model families (Table~\ref{tab:models}).
Same-family models show joint error correlations up to 60\% regardless of
prompt diversity, and single-family multi-agent setups systematically
overstate coordination gains~\cite{kim2025correlatederrors,zhang2025stopovervaluing}. If two adversarial roles share a family,
their claim sets may converge at the span level, making the exchange vacuous.

\begin{table}[t]
\caption{Model assignments. Five roles, five distinct families.
All APIs accessed at free or prototyping tier except Claude Sonnet
(Anthropic paid tier).}
\label{tab:models}
\centering
\begin{tabular}{lll}
\toprule
\textbf{Role} & \textbf{Model} & \textbf{API} \\
\midrule
Generator  & Llama 3.1 8B-Instant    & Groq       \\
Auditor    & Llama 3.1 8B-Instant    & Groq       \\
SA         & GLM-4-Flash             & Zhipu AI   \\
CD         & Mistral Small           & Mistral AI \\
Arbiter    & Cohere Command R        & Cohere     \\
           & Claude Sonnet           & Anthropic  \\
           & Llama 3.3 70B-Versatile & Groq       \\
\bottomrule
\end{tabular}
\end{table}

\begin{figure}[tb]
\centering
\resizebox{0.62\textwidth}{!}{%
\begin{tikzpicture}[
  font=\small,
  >={Stealth[length=2mm,width=1.8mm]},
  arrow/.style={->, thick, gray!50!black},
  biarrow/.style={<->, thick, gray!50!black},
  agentbox/.style={draw, rounded corners=5pt, minimum width=30mm,
                   minimum height=20mm, align=center, inner sep=5pt},
  exchbox/.style={draw=gray!40, rounded corners=4pt, fill=gray!4,
                  text width=46mm, align=left, inner sep=6pt, font=\scriptsize},
  arbbox/.style={draw=purple!50!black, rounded corners=4pt, fill=purple!8,
                 minimum width=28mm, minimum height=13mm, align=center,
                 inner sep=4pt, font=\scriptsize},
  invbox/.style={draw=black!55, rounded corners=4pt, fill=gray!6,
                 text width=88mm, align=center, inner sep=5pt},
  outbox/.style={draw=green!60!black, rounded corners=4pt, fill=green!6,
                 text width=88mm, align=center, inner sep=5pt},
]
\node[draw=gray!50, rounded corners=4pt, fill=gray!8,
      text width=88mm, align=center, inner sep=5pt] (inp) at (5.5, 0)
  {\textbf{Input record (flagged by Auditor)}\\[2pt]
   \scriptsize ``We selected candidate~B because younger workers adapt faster.''
   \quad [\texttt{delegate\,=\,true}]};
\node[agentbox, fill=red!8, draw=red!60!black] (sa) at (2.0, -2.2)
  {\textbf{Stereotype}\\[2pt]\textbf{Advocate}\\[4pt]
   \scriptsize GLM-4-Flash\\Zhipu AI};
\node[agentbox, fill=blue!8, draw=blue!60!black] (cd) at (9.0, -2.2)
  {\textbf{Context}\\[2pt]\textbf{Defender}\\[4pt]
   \scriptsize Mistral Small\\Mistral AI};
\node[exchbox] (exch) at (5.5, -4.6)
  {\textbf{Round 1}\\
   C001\,(SA): age generalisation [HIGH]\\
   D001\,(CD): qualifications [preserve]\\[4pt]
   \textbf{Round 2}\\
   C001: $0.61{\downarrow}$ \enskip D001: defended};
\node[font=\scriptsize\bfseries, text=purple!70!black] (qlabel) at (5.5, -6.3)
  {Three-family Arbiter quorum (majority vote)};
\node[arbbox] (arb1) at (2.2, -7.6)  {Arbiter 1\\\scriptsize Cohere Command~R};
\node[arbbox] (arb2) at (5.5, -7.6)  {Arbiter 2\\\scriptsize Claude Sonnet};
\node[arbbox] (arb3) at (8.8, -7.6)  {Arbiter 3\\\scriptsize Llama 3.3 70B};
\node[invbox] (inv) at (5.5, -9.3)
  {$\mathcal{I}_{\mathrm{ref}}$: every upheld claim ID has a matching instruction
   \textbf{ and }
   $\mathcal{I}_{\mathrm{sem}}$: $J(T_{\mathrm{SA}},T_{\mathrm{CD}})\leq\theta$
   (deterministic, zero LLM calls)};
\node[outbox] (out) at (5.5, -10.7)
  {\textbf{Verdict:} reframe C001, preserve D001 \qquad
   \textbf{Provenance:} every claim, vote, and instruction stored as typed record};
\draw[arrow] (inp.south) -| (sa.north);
\draw[arrow] (inp.south) -| (cd.north);
\draw[biarrow] (sa.south) |- (exch.west);
\draw[biarrow] (cd.south) |- (exch.east);
\draw[arrow] (exch.south) -- (qlabel.north);
\coordinate (barL) at (2.2, -6.6);
\coordinate (barM) at (5.5, -6.6);
\coordinate (barR) at (8.8, -6.6);
\draw[gray!50!black, thick] (qlabel.south) -- (barM);
\draw[gray!50!black, thick] (barL) -- (barR);
\draw[arrow] (barL) -- (arb1.north);
\draw[arrow] (barM) -- (arb2.north);
\draw[arrow] (barR) -- (arb3.north);
\draw[arrow] (arb1.south) -- (arb1.south |- inv.north);
\draw[arrow] (arb2.south) -- (inv.north);
\draw[arrow] (arb3.south) -- (arb3.south |- inv.north);
\draw[arrow] (inv) -- (out);
\end{tikzpicture}}
\caption{BiasGuard-D architecture. Two adversarial roles from distinct model
  families exchange typed \texttt{AtomicClaim} documents over bounded
  peer-to-peer rounds (bidirectional arrows); a three-family Arbiter quorum
  adjudicates by majority vote; a combined deterministic invariant
  ($\mathcal{I}_{\mathrm{ref}} \wedge \mathcal{I}_{\mathrm{sem}}$) validates
  the verdict at zero additional LLM cost, producing a queryable provenance
  trail.}
\label{fig:architecture}
\end{figure}
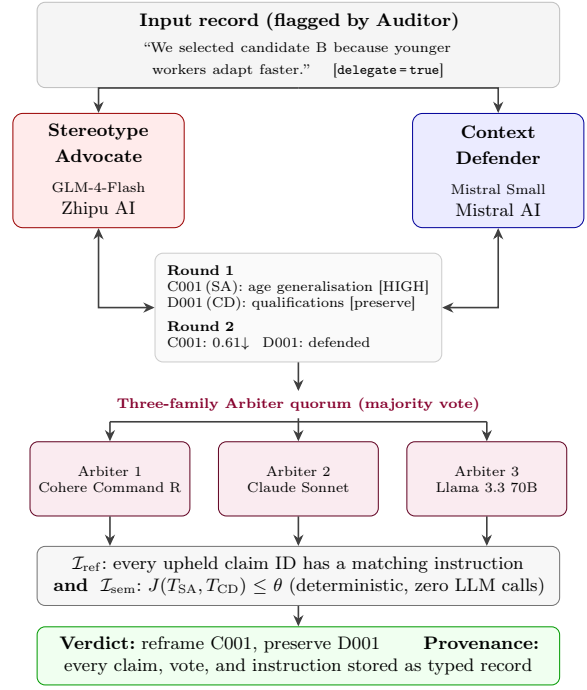

\subsection{The Combined Consistency Invariant}
\label{sec:invariant}

With typed records and stable identifiers throughout, verification reduces to
a predicate over the \texttt{DeliberationVerdict}.
It has two components, each running in $O(n)$ time with no LLM call.

\paragraph{Referential consistency ($\mathcal{I}_{\mathrm{ref}}$).}
$\mathcal{I}_{\mathrm{ref}}$ holds when every upheld SA claim maps to a
remove-or-reframe instruction and every upheld CD claim maps to a preserve
instruction, a foreign-key coverage check over the verdict's instruction
sets. A verdict that omits an instruction for any upheld claim
violates $\mathcal{I}_{\mathrm{ref}}$.

\paragraph{Semantic consistency.}
Let $T_{\mathit{SA}}$ and $T_{\mathit{CD}}$ be the span-text sets for
$U_{\mathit{SA}}$ and $U_{\mathit{CD}}$ respectively. Define
\begin{equation}
J(T_{\mathit{SA}}, T_{\mathit{CD}}) =
  \frac{|T_{\mathit{SA}} \cap T_{\mathit{CD}}|}
       {|T_{\mathit{SA}} \cup T_{\mathit{CD}}|}
\label{eq:jaccard}
\end{equation}
and let $\theta$ be a tunable threshold (pilot value 0.85, selected on a
20-item held-out BBQ Age slice and confirmed stable across the production runs).
\begin{equation}
\mathcal{I}_{\mathrm{sem}}(V) \;\equiv\;
  J(T_{\mathit{SA}}, T_{\mathit{CD}}) \leq \theta.
\label{eq:isem}
\end{equation}
In plain terms: the SA and CD must not uphold claims about near-identical text
spans, since that would instruct downstream to simultaneously remove and
preserve the same content. Referential coverage passes such verdicts; the
semantic check catches them. Together, the two checks mirror the classical
data-quality framework: $\mathcal{I}_{\mathrm{ref}}$ is a foreign-key check,
$\mathcal{I}_{\mathrm{sem}}$ is a semantic uniqueness check over the
natural-language payload.

\section{Evaluation}
\label{sec:evaluation}

The experimental study tests each question against a specific condition:
whether the semantic component adds value is tested by comparing
$\mathcal{I}_{\mathrm{ref}}$-only against the combined invariant across all
datasets; whether stored records carry diagnostic value is tested by reading
coordination metrics against output-level results on the same items; and
whether the deliberation layer transfers is tested by running it on HaluEval
QA with role-framing prompt changes and a cross-family second auditor at the
triage gate.

\subsection{Experimental Setup}
\label{sec:setup}

Both benchmarks label constraint failures on fixed-schema records, so both
fit naturally into the same typed-contract pipeline as data-quality corpora.

\textbf{BBQ Age} (primary stereotype condition, $n{=}115$) uses ambiguous-arm
items from the Bias Benchmark for QA Age category, where the context does not
entail any demographic option and a constraint-satisfying record should select
the neutral answer. We run two configurations: the full peer-to-peer
heterogeneous pipeline (P2P, $n{=}50$) used wherever deliberation-specific
signals such as stability rate or mean rounds are reported, and a single-round
mediated ablation (SR-Med, $n{=}115$) used to bound the invariant-satisfaction
estimate when the iterative loop is removed.

\textbf{BBQ Disability} (within-domain transfer, $n{=}50$) exercises the same
schema on a different demographic axis with two configurations: full P2P and
SR-Med. This pair directly isolates the contribution of the iterative loop on
a held-out stereotype category.

\textbf{HaluEval QA} (cross-domain transfer, $2{\times}n{=}50$) consists of
Wikipedia-grounded items each carrying a question, a right answer, and a
synthesised hallucinated answer. We run the same pipeline twice on disjoint
slices: once on the hallucinated answer (the data-quality framing) and once on
the right answer (a precision control measuring whether the pipeline damages
already-correct records).

Three classes of metric carry the analysis. \textbf{Data-quality metrics}:
satisfaction rate of $\mathcal{I}_{\mathrm{ref}} \wedge
\mathcal{I}_{\mathrm{sem}}$ and each component separately, and the resulting
quarantine rate. These are computed as set operations on stored records with
no model call, making them usable as production monitoring signals. \textbf{Coordination metrics}: P2P stability
rate (fraction of items whose claim-id sets reach a fixed point before the
five-round cap), mean deliberation rounds, and inter-Arbiter agreement.
\textbf{Application-level outcomes}: for BBQ, the stereotyped-option pick rate
(lower is better); for HaluEval, the hallucination-removal rate and right-answer
match rate on hallucinated inputs, and the preservation rate on correct inputs.

\subsection{Baseline Comparison and the Cost of Suppression}
\label{sec:baselines}

Table~\ref{tab:baselines} places BiasGuard-D alongside three simpler
interventions on BBQ Age ($n{=}115$): zero-shot (ZS), a stereotype-aware system
prompt (SDP), and chain-of-thought (CoT).

\begin{table}[!ht]
\caption{BBQ Age stereotype pick rate under four conditions ($n{=}115$
ambiguous-arm items). ZS = zero-shot; SDP = stereotype-aware system prompt;
CoT = chain-of-thought. $^{*}$CoT routes 108/115 items to the neutral option
regardless of content. $^{\dagger}$SDP raises neutral responses from 30.4\%
(ZS) to 65.2\%; within-item shifts show zero classification changes,
suggesting a model-calibration effect rather than item-level remediation.}
\label{tab:baselines}
\centering
\small
\begin{tabular}{l c @{\hspace{8mm}} r @{\hspace{8mm}} l}
\toprule
\textbf{Condition} & \textbf{Stereotype rate} & \textbf{$\Delta$ vs.\ ZS} & \textbf{Note} \\
\midrule
ZS              & 51.3\% & ---        & Generator baseline \\
SDP             & 24.4\% & $-$26.9pp  & Prompt-only, no audit trail$^{\dagger}$ \\
CoT             &    0\% & $-$51.3pp  & 94\% unknown collapse$^{*}$ \\
BiasGuard-D     & 40.0\% & $-$11.3pp  & Typed invariant + provenance \\
\bottomrule
\end{tabular}
\end{table}

CoT's 0\% stereotype rate is the most instructive number in Table~\ref{tab:baselines},
and not in CoT's favour. It is achieved not by reasoning about each item but
by routing 108 of 115 answers to ``can't be determined'' irrespective of the
question. The model has learned to suppress rather than reason. SDP shows the
same pattern on a lesser scale: neutral responses rise from 30.4\% under ZS to
65.2\% under SDP, and within-item shift analysis finds zero classification
changes: every item that was stereotyped under ZS remained stereotyped under
SDP, and vice versa. The aggregate rate difference between ZS and SDP reflects
a model-calibration shift, not item-level remediation; and like CoT, SDP
leaves no record of which spans were judged problematic or why.
BiasGuard-D produces a smaller raw reduction ($-$11.3pp), but every verdict is
a typed, diagnosable record; CoT and SDP ``succeed'' by suppression, leaving
nothing to audit and no way to tell whether it is principled or reflexive.

\subsection{Invariant Analysis: Catching What Referential Coverage Misses}
\label{sec:invariant-analysis}

Across every dataset, every fallback is triggered by $\mathcal{I}_{\mathrm{sem}}$
failing on a record that fully satisfies $\mathcal{I}_{\mathrm{ref}}$ (the FB\%
column of Table~\ref{tab:invariant} equals the $\mathcal{I}_{\mathrm{sem}}$ violation
count to within zero in every row). This is direct evidence that the semantic check adds value: there is a
recurring class of verdict records $V$ where $\mathcal{I}_{\mathrm{ref}}(V) = \top$
and $\mathcal{I}_{\mathrm{sem}}(V) = \bot$.

The size of the violating class depends on task and deliberation mode. The
clearest comparison is BBQ Disability, where the two configurations share
identical 36 delegated items: under SR-Med the semantic violation rate is 44.4\%
(Wilson 95\% CI $[29.5,\,60.4]$), under P2P it falls to 16.7\%
($[7.9,\,31.9]$), and the intervals do not overlap. Iterative deliberation is the only difference between the two runs, so this
gap shows the loop is resolving the contradictions the semantic check is
designed to catch. BBQ Age under SR-Med has only a 2.6\% violation rate, not
because mediated deliberation resolved anything, but because the Age items
rarely produce near-duplicate SA/CD claims to begin with. The semantic invariant
is not a theoretical safeguard; it is what separates a resolved contradiction
from a suppressed one, and its failure rate varies predictably with the task.

The $\theta{=}0.85$ threshold is robust to variation (Table~\ref{tab:theta}):
the Jaccard distribution is bimodal with all violating items at $J{=}1.0$
and non-violating items below $J{=}0.5$, so quarantine rates are identical
across $[0.70,\,0.85]$ and shift by only 1.3pp at $\theta{=}0.90$
(one BBQ Age item at $J{=}0.875$). HaluEval's maximum Jaccard is $0.69$,
below any tested threshold.

\begin{table}[!ht]
\caption{$\mathcal{I}_{\mathrm{sem}}$ quarantine rate (FB\%) across Jaccard
thresholds; production value $\theta{=}0.85$ in bold. All violations fall
at $J{=}1.0$ except one BBQ Age item ($J{=}0.875$), the only
threshold-sensitive case.}
\label{tab:theta}
\centering
\small
\begin{tabular}{l c c c c}
\toprule
\textbf{Condition} & $\theta{=}0.70$ & $\theta{=}0.80$
  & $\boldsymbol{\theta{=}0.85}$ & $\theta{=}0.90$ \\
\midrule
BBQ Age, P2P        & 25.0 & 25.0 & \textbf{25.0} & 23.7 \\
BBQ Disability, P2P & 16.7 & 16.7 & \textbf{16.7} & 16.7 \\
HaluEval QA, hall.  &  0.0 &  0.0 & \textbf{ 0.0} &  0.0 \\
\bottomrule
\end{tabular}
\end{table}

\subsection{Schema-Level Coordination as Diagnostic Signal}
\label{sec:coordination}

On BBQ Disability, the SR-Med and P2P configurations reach very different
invariant-satisfaction rates (55.6\% versus 83.3\%, Wilson 95\% CIs
$[39.6,\,70.5]$ vs.\ $[68.1,\,92.1]$) at identical inter-Arbiter agreement
(0.57). Agreement alone does not separate them; the difference is that the
iterative loop reduces internal claim-set overlap across rounds before the
Arbiters see them. The deliberation mode leaves a signature in stored records
that is readable independently of the final output, which is the diagnostic
value the coordination metrics are designed to surface.

The HaluEval condition makes the converse point. Inter-Arbiter agreement is
0.98 on the hallucinated inputs, far above the 0.49--0.57 range on BBQ.
This is not a calibration effect; it reflects that an objective grounding
task gives heterogeneous Arbiters something concrete to converge on.
Subjective stereotype categories leave room for genuine disagreement, and the
typed records preserve that disagreement rather than smooth it over. The
coordination metrics thus carry diagnostic value beyond pick rate or removal
rate, telling an operator whether deliberation is converging or simply
running out the clock.

\subsection{Cross-Domain Transfer}
\label{sec:crossdomain}

The HaluEval rows of Tables~\ref{tab:invariant} and~\ref{tab:coordination} answer the cross-domain question directly. On hallucinated
inputs the hallucination-removal rate is 100\% (43/43 confirmed cases); on the
same items the right-answer match rate rises from 12.2\% before remediation to
61.2\% after. The remaining outputs are conservative hedges, which is the correct behaviour
when the reference passage does not fully entail the right answer;
asserting it anyway would introduce facts not in the source.

The precision control on already-correct inputs tells the symmetric story.
Of 50 right-answer records the dual auditor delegates 24, of which 11 are
preserved verbatim and 13 become hedges, giving an overall preservation rate
of 71.4\%. Of the 24 false delegations, 19 fire because one auditor flags
while the other passes; only 2 are consensus mis-flags. Requiring both
auditors to agree before delegating would eliminate about 92\% of false
delegations on correct inputs, at the cost of routing 22\% of true
hallucinations through the pass-through path. This is the standard
recall-precision trade-off, and the architecture exposes it at the one stage
where an operator can tune it without touching the data model, quorum, or
invariant.

The deliberation layer and invariant thus handle both a subjective social-bias
task and an objective factual-grounding task with only role-framing changes,
the portability claim the architecture was designed to support. The triage gate
is the one component that required domain adaptation: a cross-family second
auditor is needed at the gate to maintain recall when the generator's own family
produces the hallucination it is asked to detect.

\begin{table}[t]
\caption{Invariant metrics across conditions. Inv.\% = combined-invariant
satisfaction rate on delegated items (Wilson 95\% CI in brackets);
FB\% = quarantine rate (equals $\mathcal{I}_{\mathrm{sem}}$ violation count).
P2P = full peer-to-peer; SR-Med = single-round mediated (no iterative loop).
$^{\dagger}$No $\mathcal{I}_{\mathrm{sem}}$ violations; the 4.2pp gap is one
item whose deliberation agent emitted malformed JSON on both retry attempts
(traceable to conflicting auditor opinions on a date-format mismatch)
and is excluded from FB\% as a pipeline execution failure rather than a
semantic contradiction. The stored error record identifies the item, the
failure point, and the fallback output, illustrating the provenance trail
the architecture is designed to provide.}
\label{tab:invariant}
\centering
\small
\begin{tabular}{l c c c c c}
\toprule
\textbf{Dataset / cond.} & \textbf{$n$} & \textbf{Deleg.}
  & \textbf{Inv.\%} & \textbf{[95\% CI]} & \textbf{FB\%} \\
\midrule
BBQ Age, P2P           & 115 & 76 & 75.0 & [64.2,\,83.4] & 25.0 \\
BBQ Disability, P2P    &  50 & 36 & 83.3 & [68.1,\,92.1] & 16.7 \\
BBQ Disability, SR-Med &  50 & 36 & 55.6 & [39.6,\,70.5] & 44.4 \\
HaluEval QA, hall.     &  50 & 49 & 100  & [92.7,\,100]  &  0.0 \\
HaluEval QA, right     &  50 & 24 & 95.8 & [79.8,\,99.3] &  0.0$^{\dagger}$ \\
\bottomrule
\end{tabular}
\end{table}

\begin{table}[t]
\caption{Coordination and outcome metrics across conditions.
Arb.\ Agr.\ = mean inter-Arbiter per-claim agreement;
Stab.\% = P2P fixed-point rate; Rds = mean rounds before stability;
$\Delta$ / rate = application-level outcome: percentage-point reduction
in stereotype pick rate (BBQ), pp recovery in right-answer match
(HaluEval hall.), or absolute preservation rate (HaluEval right$^{\ddagger}$).
$^{*}$Right-answer match $12.2\% \rightarrow 61.2\%$; hallucination
removal $43/43 = 100\%$.
$^{\ddagger}$Preservation rate on already-correct inputs (precision control).}
\label{tab:coordination}
\centering
\small
\begin{tabular}{l @{\hspace{8mm}} c @{\hspace{8mm}} c @{\hspace{6mm}} c @{\hspace{8mm}} c}
\toprule
\textbf{Dataset / cond.} & \textbf{Arb.\ Agr.}
  & \textbf{Stab.\%} & \textbf{Rds} & \textbf{$\Delta$ / rate} \\
\midrule
BBQ Age, P2P           & 0.82 & ---  & ---  & $-13.0$pp \\
BBQ Disability, P2P    & 0.57 & 100  & 2.39 & $-22.0$pp \\
BBQ Disability, SR-Med & 0.57 & ---  & 1.0  & $-8.0$pp  \\
HaluEval QA, hall.     & 0.98 & 93.9 & 2.24 & $+49.0$pp$^{*}$ \\
HaluEval QA, right     & 0.87 & 79.2 & 2.30 & $71.4\%^{\ddagger}$ \\
\bottomrule
\end{tabular}
\end{table}

\section{Discussion and Future Work}
\label{sec:discussion}

Three threads extend beyond the experimental numbers: where the architecture
matters in practice, what the cross-domain transfer demonstrated, and where
the current design falls short.

\subsection{Deployment Use Cases and Monitoring Signals}
\label{sec:usecases}

The conflicted-record problem arises wherever LLM-generated text is treated
as a data asset. In \emph{content moderation}, advocate roles become a
removal-advocate and a preservation-advocate; the invariant catches the case
where both ask for incompatible actions on overlapping spans. In
\emph{knowledge-base ingestion}, a grounding-advocate and a
contextualisation-advocate produce claims over extracted facts, and the
invariant prevents a verdict from simultaneously asserting and retracting the
same fact. In \emph{privacy-versus-utility data release}, a privacy-advocate
and a utility-advocate flag the same spans for opposite reasons; the
architecture surfaces this conflict as typed records a data steward can audit,
rather than burying it in a classifier score.

A high FB\% does not signal pipeline failure: it signals a genuine conflict
the deliberation could not resolve, where a conservative fallback is correct
and a forced verdict would instruct the system to simultaneously remove and
preserve the same span. On BBQ Disability the iterative loop resolves roughly
two-thirds of such conflicts (quarantine drops from 44.4\% under SR-Med to
16.7\% under P2P).

Agreement rate, semantic violation rate, and claim-count asymmetry can each
be tracked over rolling windows with no additional classifier, aligning with
inspectability requirements under NIST AI RMF~\cite{nist2023airmf} and the
EU AI Act~\cite{euaiact2024}.

\subsection{Limitations}
\label{sec:limitations}

Two limitations bear on the evidence. The LLM-as-judge step that produced the HaluEval hallucination-removal
labels is itself subject to the same biases the architecture is designed
to catch. We mitigate by giving the judge a strict comparison prompt with no
reference to intermediate records, but a fully independent human judge would
eliminate this concern entirely.

One limitation bears on the auditor design. The HaluEval condition uses a
two-auditor gate with \texttt{OR} delegation; 19 of 24 false delegations on
correct inputs fire because one auditor flags while the other passes. The
delegation strategy is a tunable threshold that does not touch the data
model, quorum, or invariant; the trade-off is examined in future work.

One further limitation bears on configuration. The Jaccard threshold has not
been tested on paraphrase-rich corpora; Table~\ref{tab:theta} shows the
quarantine rate is insensitive to $\theta$ across $[0.70,\,0.90]$ on the
current datasets, but a denser paraphrase space could push items into the
borderline band and raise the false-positive rate.
Most roles use free or prototyping tiers deliberately, to stress-test the
architecture under realistic resource constraints. The exception is
Claude Sonnet as one Arbiter: ablations replacing it with free-tier
alternatives (Qwen3-32B, Mistral Large) produced substantially lower
hallucination removal rates on HaluEval, showing that Arbiter capability
is a genuine dependency on objective grounding tasks. The BBQ stereotype
results, which do not rely on this slot, are unaffected. Identifying a fully free-tier Arbiter that matches Claude-level performance
on HaluEval and improving end-to-end latency are left as future work.

\subsection{Future Work}
\label{sec:futurework}

Four directions extend the current system. First, auditor configuration:
switching from \texttt{OR} to \texttt{AND} delegation, so that both auditors
must agree before escalating, eliminates roughly 92\% of false
delegations on correct inputs without additional inference cost, at the
cost of routing some true hallucinations through the pass-through path;
evaluating this trade-off systematically across domain types is the
nearest-term engineering extension. A three-auditor majority vote would
break most ties but multiplies cost across the full record stream, making
it less practical at deployment scale. Second, the Jaccard threshold has
been validated only on the current BBQ and HaluEval distributions;
paraphrase-rich corpora where surface overlap is low but semantic overlap
is high may require a dense-embedding similarity check to replace or
supplement it. Third, the portability claim currently rests on two
domains; extending to content moderation, knowledge-base ingestion, and
privacy-versus-utility data release would establish whether only
role-framing changes suffice at the triage gate, or whether the invariant
itself needs domain-specific extension. Fourth, replacing the LLM-as-judge
evaluation labels with independent human annotation would strengthen the
HaluEval evidence base, and identifying a fully free-tier Arbiter matching
Claude Sonnet-level performance on objective hallucination tasks remains an
open engineering problem.

\section{Conclusion}
\label{sec:conclusion}

LLM-generated records increasingly populate the enterprise systems where
output quality and inspectability are operational requirements. When a single
record both satisfies and violates a quality rule, no static validation layer
can resolve the conflict: the tension is semantic, and resolving it requires
deliberation between parties with opposing biases, not a threshold check. The
question this paper asked is whether a deterministic check over structured
inter-agent records can catch the quality defects that referential checks alone
miss, and whether those records carry enough information to audit pipeline
quality without calling an additional model.

We addressed that question by redesigning the data path: two adversarial roles
from distinct model families exchange typed claims over peer-to-peer rounds, a
three-family quorum adjudicates, and a combined invariant validates the result.
The semantic component fires on verdicts that referential checks miss; removing
the iterative loop drops invariant satisfaction by nearly 30 percentage points
on the BBQ Disability ablation; and the deliberation layer, data model, and
invariant transfer from stereotype to hallucination remediation with only
role-framing changes, raising right-answer recovery by 49 points, with the
triage gate as the one domain-sensitive component. Every claim, vote, and
instruction is stored as a typed record that an operator can query without
re-invoking any model.

The contribution is a portable template. Any task where a record both satisfies
and violates a quality rule can instantiate the data model by defining
domain-appropriate claim types, assigning advocate roles with opposing biases,
and keeping the same combined invariant. More broadly, data-management discipline can now extend, with
auditable runtime guarantees, to records produced by LLMs, the records that
increasingly populate the enterprise systems where inspectability is becoming
a deployment requirement.

{\setlength{\itemsep}{0pt}\setlength{\parsep}{0pt}%
\bibliographystyle{splncs04}
\bibliography{references}}

\end{document}